\documentclass[aps, prapplied, reprint, superscriptaddress]{revtex4-1}
\usepackage{latexsym}
\usepackage{amsthm}
\usepackage{amssymb}
\usepackage{amsmath}
\usepackage[all]{xy}
\usepackage{float}
\usepackage{graphicx}
\usepackage{subfigure}
\usepackage{dcolumn}
\usepackage{bm}
\usepackage{bbm}
\usepackage[usenames, dvipsnames]{xcolor}
\usepackage{amsfonts}
\usepackage{cases}
\usepackage[a4paper, pagebackref=false, colorlinks=true,
linkcolor=blue, citecolor=blue,
pdfauthor={ },
pdftitle={ },
pdfsubject={ },
pdfkeywords={ }]{hyperref}

\begin{document}

\title{Manipulation of a Micro-object Using Topological Hydrodynamic Tweezers}

\author{Peiran Yin}
\thanks{These authors contributed equally to this paper and are joint first
	authors.}
\affiliation{CAS Key Laboratory of Microscale Magnetic Resonance and Department of Modern Physics, University of Science and Technology of China, Hefei 230026, China}
\affiliation{Synergetic Innovation Center of Quantum Information and Quantum Physics, University of Science and Technology of China, Hefei 230026, China}

\author{Rui Li}
\thanks{These authors contributed equally to this paper and are joint first
	authors.}
\affiliation{CAS Key Laboratory of Microscale Magnetic Resonance and Department of Modern Physics, University of Science and Technology of China, Hefei 230026, China}
\affiliation{Synergetic Innovation Center of Quantum Information and Quantum Physics, University of Science and Technology of China, Hefei 230026, China}
\affiliation{Hefei National Laboratory for Physical Sciences at the Microscale, University of Science and Technology of China, Hefei 230026, China}
\author{Zizhe Wang}
\author{Shaochun Lin}
\author{Tian Tian}
\author{Liang Zhang}
\author{Longhao Wu}
\author{Jie Zhao}
\author{Changkui Duan}
\affiliation{CAS Key Laboratory of Microscale Magnetic Resonance and Department of Modern Physics, University of Science and Technology of China, Hefei 230026, China}
\affiliation{Synergetic Innovation Center of Quantum Information and Quantum Physics, University of Science and Technology of China, Hefei 230026, China}
\author{Pu Huang}
\thanks{Corresponding author: \href{mailto:hp@nju.edu.cn}{hp@nju.edu.cn}}
\affiliation{National Laboratory of Solid State Microstructures and Department of Physics, Nanjing University, Nanjing, 210093, China}
\author{Jiangfeng Du}
\thanks{Corresponding author: \href{mailto:djf@ustc.edu.cn}{djf@ustc.edu.cn}}
\affiliation{CAS Key Laboratory of Microscale Magnetic Resonance and Department of Modern Physics, University of Science and Technology of China, Hefei 230026, China}
\affiliation{Synergetic Innovation Center of Quantum Information and Quantum Physics, University of Science and Technology of China, Hefei 230026, China}
\affiliation{Hefei National Laboratory for Physical Sciences at the Microscale, University of Science and Technology of China, Hefei 230026, China}
\begin{abstract}

Manipulating micro-scale object plays paramount roles in a wide range of fundamental researches and applications. At micro-scale, various methods have been developed in the past decades, including optical, electric, magnetic, aerodynamic and acoustic methods. However, these non-contact forces are susceptible to external disturbance, and so finding a way to make micro-scale object manipulation immune to external perturbations is challenging and remains elusive. Here we demonstrate a method based on new trapping mechanism to manipulate micro-scale object in a gas flow at ambient conditions. We first show that the micro-droplet is entrapped into a trapping ring constructed by a particular toroidal vortex. The vortex works as tweezers to control the position of the micro-droplet. We then show that the micro-droplet can be transported along the trapping ring. By virtue of the topological character of the gas flow, the transport path is able to bypass external strong perturbations automatically. We further demonstrate a topological transfer process of the micro-droplet between two hydrodynamic tweezers. Our method provides an integrated toolbox to manipulate a micro-scale object, with an intrinsic mechanism that protects the target object from external disturbances.

\end{abstract}  
\maketitle

\section{INTRODUCTION}
Manipulating micro-scale object is always a challenging issue in various fundamental researches and applications. Because methods based on direct mechanical contact fail at micro-scale, various non-contact methods have been developed in the past decades, including optical \cite{Optical_tweezer}, electric \cite{Paul_trap}, magnetic \cite{magnetic_trap}, aerodynamic\cite{aerodynamic} and acoustic \cite{Acoustic_trap} methods. These methods have found wide applications in physics \cite{Cold_atom, Levitated_optomechanics}, chemistry \cite{chemistry}, biology \cite{Cell, DNA} and information science \cite{information1, information2}. However, due to the fragility of non-contact force, the manipulation of micro-scale object is often susceptible to external disturbance.

Topological phenomena have been observed in various physics systems recently, involving charge \cite{quantum_hall_effects} and spin transport \cite{topological_insulator1, topological_insulator2}, optical photons \cite{Optical_1, Optical_2, Optical_3, Optical_4}, microwave \cite{Microwave_1, Microwave_2}, mechanical vibrations \cite{Mech_1} and acoustics \cite{acoustic_2}. One topological phenomena's character is that, the global transport process is immune to local disturbance \cite{immune}.

In fluid dynamics, the vorticity $\bm{\Omega}$ is a pseudovector field that describes the vortex field, and is defined as the curl of the flow velocity $\bm{\upsilon}$, i.e., $\bm{\Omega} = \nabla \times \bm{\bm{\upsilon}}$. For an incompressible fluid, vorticity has the special property that its field lines never cross. This indicates an important topological property that knotted or linked loops in the fluid will remain knotted or linked even in the presence of viscosity or external local perturbations \cite{fluid1, fluid2}. Besides fundamental interest, whether such topological properties have any practical application remains to be explored.

Here, we generate a particular toroidal vortex using a tapering micro-nozzle, which works as tweezers to manipulate the micro-droplet at ambient conditions. In this trapping mechanism, the particle released in gas flow is entrapped in the vortex core under the action of Stokes drag force from the vortex flow and a restoring force in gravity direction. And by taking advantage of the non-local topological property of the vortex flow field, we have further demonstrated the transport and transfer of the object in a controllable way that is intrinsically protected from external disturbance.

\section{CONCEPT OF THE HYDRODYNAMIC TWEEZERS}
For a micro-particle released in a non-uniform flow,  determining the drag and lift forces  arising  from both inertia and unsteady effects at low Reynolds number is a long-standing open question in fluid dynamics \cite{force1,force2,force3}. Intuitively, when a micro-object is released in a swirling fluid flow, it is predicted to spiral towards the equilibrium point close to the center of vortex core. Actually, it is verified in our experiment that, at a low Reynolds number around 0.001, a heavy micro-particle in the particular toroidal vortex is entrapped into a trapping ring spirally. And we can manually pull the micro-particle to transport along the ring.

Before building practical hydrodynamic tweezers, three issues need to be addressed. First, the size of object to be trapped is dependent on the characteristic scale of fluid flow, and so it is possible in principle to realize the control of an arbitrarily small object. However, as the system scales down, the condition of continuous fluid would fail and the molecular behaviors would emerge. The typical value of mean free path is tens of nanometers for gas phase flow in practical realization, and is sub-nanometer when liquid phase fluid is used. The second issue is the fluctuation of the fluid field. As the fluid flows past an object, the object also exerts a back action onto the fluid. This effect is characterized by the Reynolds number ($Re$). When $Re$ is large enough, turbulence occurs and the system becomes unstable. For micro-objects, however, the typical $Re$ is usually small enough to avoid turbulence. The third issue is related to the viscosity. As a vortex in a non-ideal fluid will eventually spread and damp out, suitable external pump is necessary to maintain the main structure of the vortex.


The scheme of our hydrodynamic tweezers is shown in Fig.~1(a). The gas flows from a tapering micro-nozzle and generates a toroidal vortex downstream. A micro-scale particle set around the vortex tube feels a trapping force $\bm{F}_{\textrm{T}}$, which attracts the object towards vortex center that forms a ring in the $x$-$y$ plane. So such a particle can be trapped on the ring steadily with this hydrodynamic tweezers.

 The generated flow field is symmetric around the $z$ axis. So the flow patterns can be clearly demonstrated by the streamlines \cite{streamline} in the $y$-$z$ plane as shown in Fig.~1(b). The stream function is defined by
 \begin{align}
 \Psi =  \int_{0}^{y} v_zy \mathrm{d}y,
 \end{align}
 where $v_z$ is the magnitude of velocity of the gas flow in $z$ direction. The stream function is obtained by a numerical simulation of the flow field which is described in detail in Sec. \uppercase\expandafter{\romannumeral1} of the supplemental material \cite{supplementary}. As indicated in Fig.~1(b), the vortex core is formed over one nozzle diameter length downstream. The flow around vortex core can be approximately described as a forced vortex in the $y$-$z$ plane with vortex center at the vortex core. And the angular velocity $\omega$ is around $2\pi\times16$ Hz. The velocity and vorticity fields in the cross section shown as dotted box in Fig.~1(b) are indicated by blue and yellow marks in Fig.~1(c). The vorticity $\bm{\Omega}$ is invariant around the vortex center with $\Omega=2\omega$. With distance from vortex center no more than $50$ $\mu$m, the maximum velocity deviation from the forced vortex model is within $20\%$.
 \begin{figure}
 	\centering
 	\includegraphics[width=0.95\columnwidth]{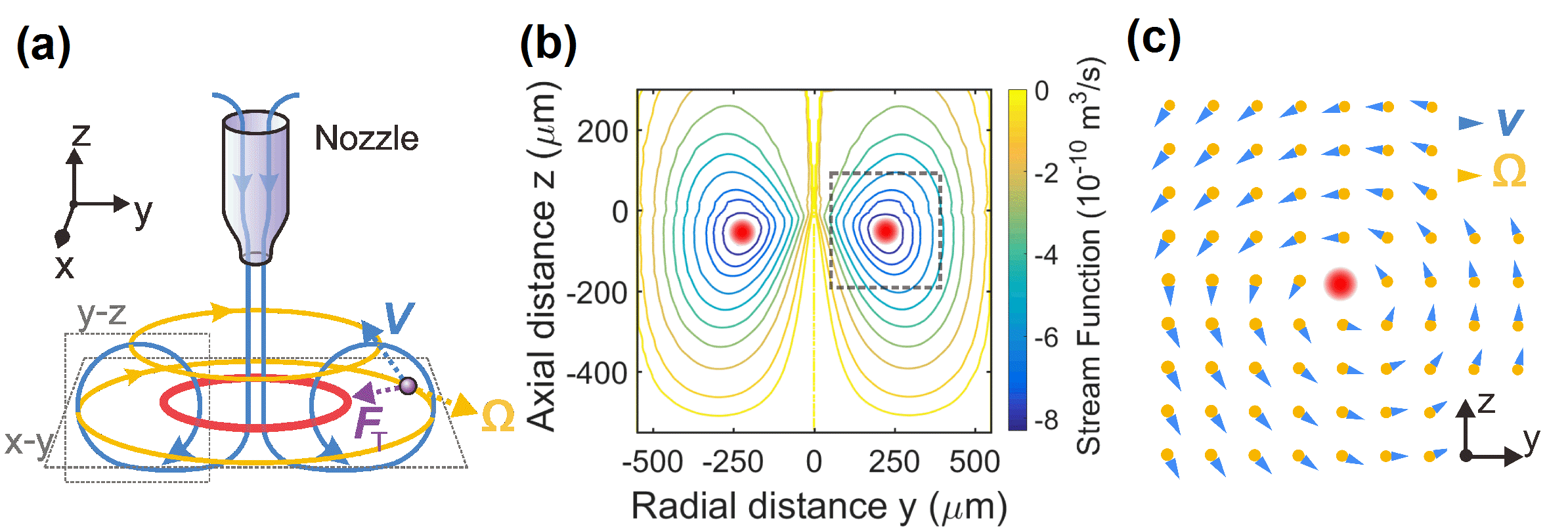}
 	\caption{ Concept of the hydrodynamic tweezers.
 		(a) Schematic plot of the system. Gas flow indicated as blue field lines come from a micro-nozzle and generate a vorticity field $\bm{\Omega}$, indicated as yellow field lines. An object is indicated as a purple sphere, which spins with the vortex field. This object feels a trapping force $\bm{F}_\textrm{T}$, which pulls it towards the trapping ring shown as a red circle. The object can move along the ring, and is stably trapped in the ring.
 		(b) Streamlines of the fluid flow in $y$-$z$ plane. The red dot marks the trapping point.
 		(c) Numerical simulation of the velocity and vorticity fields in the cross section shown as dotted box in (b). They are indicated by blue arrows and yellow dots respectively. The size of marks is proportional to the field strength and the region is about $100$ $\mu$m by $100$ $\mu$m. In the simulation, the nozzle diameter is $20$ $\mu$m, and the maximum velocity at nozzle outlet is $3.2$ ms$^{-1}$ (see Sec. \uppercase\expandafter{\romannumeral1} of the supplemental material \cite{supplementary}).
 	}\label{dips}
 \end{figure}

  For a particle released near the toroidal vortex at low Reynolds number, it feels drag and lift forces from the flow \cite{force1,force2,force3}. With the main concern being about the trapping process near the vortex core, it is reasonable to approximately treat the flow in the vicinity of the vortex core in one cross section (i.e. the $y$-$z$ plane) as a forced vortex. Then the trapping dynamics can be further studied analytically.
  
  \section{TRAPPING THEORY}
  The Reynolds number of the flow is $Re= d {\upsilon} \rho/\mu$, where $d$ is the diameter of the nozzle, $\upsilon$ is the magnitude of velocity of fluid flow, $\rho$ and $\mu$ are the fluid's density and viscosity coefficient respectively. For the micro-nozzle in our scheme, $Re$ is much less than $10$, which is far from turbulent condition. Similarly, the particle Reynolds number $Re_p= a {\upsilon_s} \rho/\mu$  is around 0.001. Here, $a$ is the radius of particle and ${\upsilon_s}$ is the magnitude of relative velocity of particle and fluid measured on the streamline through the center of particle (i.e. $\bm{v}_\textrm{s}=\bm{v}_\textrm{p}-\bm{v}$). 
  
  In our setup, a liquid micro-scale droplet is released in a gas flow at ambient conditions. So the buoyancy and the classical pressure gradient of the unperturbed flow can be ignored comparing with Stokes drag force. The force analysis of the droplet in fluid flow is shown in Fig~2(a). The gravity of the droplet is balanced by the diamagnetic force generated by an array of magnets (see Sec. \uppercase\expandafter{\romannumeral2} of the supplemental material \cite{supplementary} for more detail). Therefore, the droplet feels a restoring force $\bm{F}_{\textrm{m}}=-k\bm{z}$  in the $z$ direction near the equilibrium position. In the limit of $ 1 \gg T_a^{1/2} \gg Re_p$, where $T_a = a^2\omega \rho/\mu$ is the Taylor number, the correction to the Stokes drag force $\bm{F}_\textrm{D}$ is in fact the time-dependent history force $\bm{F}_\textrm{H}$ on the particle \cite{force2,force3}, which is of the order $\sim T_a^{1/2} 6\pi\mu a\bm{v}_\textrm{s}$. In our scheme, $T_a^{1/2}$ is around 0.01, so the rate of approaching to the trapping point is modified only slightly by the small history force. As a consequence, the key forces driving this heavy liquid droplet to the stable trapping point are the Stokes drag force from such a vortex and a restoring force $\bm{F}_{\textrm{m}}$ just in the $z$ direction. So the trapping force can be briefly expressed as
   \begin{align}
   \bm{F}_\textrm{T}=\bm{F}_\textrm{D}+\bm{F}_\textrm{m}=-6\pi\mu a\bm{v}_\textrm{s}-k\bm{z}.
   \end{align}
   Hence, a droplet under the stable balance of an external force $\bm{F}_\textrm{ext}$ and the trapping force stays stably at one position away from vortex center. Increasing flow velocity can strengthen the trapping force, so the tweezers are stronger to overcome external interactions. This adjustable trapping force lets the hydrodynamic tweezers have more practical applications.
    \begin{figure}
   	\centering
   	\includegraphics[width=0.95\columnwidth]{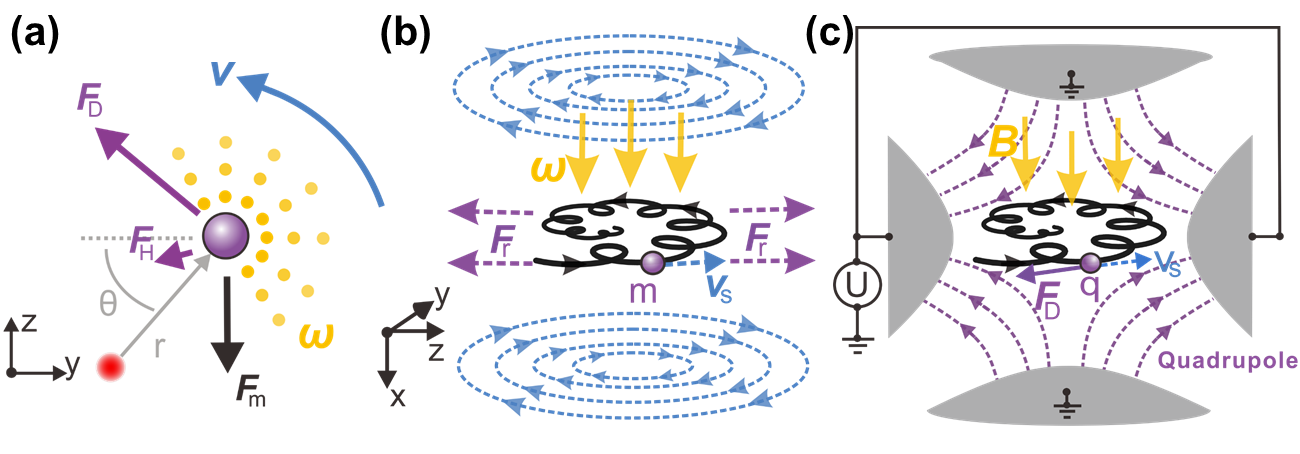}
   	\caption{ Trapping dynamics.
   		(a) Force analysis of the droplet in $y$-$z$ plane. The red dot indicates the vortex center, and the purple sphere represents the droplet. The angular velocity $\omega$ is constant.
   		(b) A sketch of dynamics of the object in rotating reference frame. 
   		(c) A sketch of dynamics of the object in a Penning trap with gas damping.
   	}\label{dips}
   \end{figure}

To further analyze the dynamic process of trapping, we choose the rotating reference frame of angular velocity of $\omega\hat{\bm{x}}$, with the origin point at the vortex core, as shown in Fig~2(b). Then the particle motion equation can be written as
\begin{equation}
\begin{split}
{\left(\begin{array}{cccc} 
\ddot{y}\\\ddot{z} 
\end{array}\right)}=
(\omega^2-&\frac{k}{2m}){\left(\begin{array}{cccc} 
	y\\z 
	\end{array}\right)}
+2\omega{\left(\begin{array}{cccc} 
	\dot{z}\\-\dot{y} 
	\end{array}\right)}
-\frac{6\pi\mu a}{m}{\left(\begin{array}{cccc} 
	\dot{y}\\\dot{z} 
	\end{array}\right)}\\
&-\frac{k}{2m}{\left(\begin{array}{cccc} 
	-\cos(2\omega t)&\sin(2\omega t)\\ \sin(2\omega t)&\cos(2\omega t)
	\end{array}\right)}{\left(\begin{array}{cccc} 
	y\\z 
	\end{array}\right)}. 
\end{split}
\end{equation}
It is noted that now $(y,z)$ is the position of the particle in the rotating frame. The first two terms in the right-hand side (RHS) of Eq.~(3) describe the radial motion in a Penning trap \cite{penning} shown in Fig~2(c). 

The first term in RHS of Eq.~(3) represents the radial repulsive force $\bm{F}_\textrm{r}$ corresponding to the electrostatic potential of a Penning trap, and the second term in RHS of Eq.~(3) represents the Coriolis force corresponding to the Lorentz force of a Penning trap. Without gas damping, the perpetual radial motion of a trapped particle in Penning trap can be resolved into a fast modified cyclotron motion with angular velocity $\omega_\textrm{c}'$ and a slow magnetron motion with angular velocity $\omega_\textrm{m}$. Correspondingly, we have:
 \begin{equation}
 \begin{split} 
\omega_\textrm{c}'&=\omega+\sqrt{\frac{k}{2m}}\\
\omega_\textrm{m}&=\omega-\sqrt{\frac{k}{2m}}.
\end{split}
\end{equation}
 Now, we consider the buffer gas cooling of a penning trap \cite{cooling}. The damping effect refers to the third term in RHS of Eq.~(3). Under the damping effect, the energy of cyclotron motion will decrease, and its radius shrinks. But the magnetron motion has a maximum total energy at the center of the trap because of its negative sign of energy. The buffer gas damping will cause the particle to move down the magnetron energy hill to a larger radius, which means the failure of trapping particle to the center. Hence, we need an azimuthal rf quadrupole field in the trap at frequency $2\omega=\omega_\textrm{c}'+\omega_\textrm{m}$ to realize a parametric process that couples the two motions \cite{parametric}. This is just the last term in RHS of Eq.~(3). The parametric coupling can convert the magnetron motion to the modified cyclotron motion, and combined with the gas damping, will cause the magnetron radius also to decrease. Then the particle can be entrapped steadily to the center, as indicated by the trajectories in Figs~2(b) and 2(c).

Hence, the trapping mechanism of the particle can be understood as a gas damping process of a Penning trap using ($\omega_\textrm{c}'+\omega_\textrm{m}$) parametric coupling. And then the condition necessary to achieve trapping is obtained as follows:
  \begin{align}
 \omega \leq \omega_\textrm{max}=\sqrt{\frac{k_0}{2\rho}+(\frac{k_0a^2}{9\mu})^2},
 \end{align}
 where $k_0=3k/(4\pi a^3)$. When $\omega=\omega_\textrm{max}$, it is the critical condition that the parametric term offsets the damping term exactly at any moment, then the particle keeps perpetual radial motion as in an ideal classic Penning trap. Turning back to the laboratory reference frame, this critical condition is that the particle travels around the vortex center in a periodic orbit with $\omega=\omega_\textrm{max}$. An angular velocity $\omega<\omega_\textrm{max}$ makes the particle spiral into the trapping center as desired; while a larger $\omega$ leads to the escape of the particle.


\section{EXPERIMENTAL MANIPULATION OF THE MICRO-OBJECT}
\subsection{Trap}

In the experimental realization schematically shown in Fig.~3(a), we employ a micro-droplet of water as the target to be controlled by the hydrodynamic tweezers. And an array of magnets are used to offset the gravity of droplet by diamagnetic effect. A tapering glass nozzle is used to generate a particular toroidal vortex in the center region of the magnets array. We use a dosing valve to precisely control the volume flow rate of the gas under an additional pressure $P$ at the inlet. A microscope is used to observe the droplet. Our experiments are carried out at room temperature using nitrogen gas for extensive use.  Figs.~3(b), 3(c) and 3(d) show the photographs of the setup, the nozzle and the droplet suspended in the region at the center of magnets array. The region is a cylindrical cavity of radius $0.6$ mm and height $1.2$ mm. It is noted that the offset force generated by the magnets can be produced by other means, such as by electric forces or dielectric forces, depending on the target material.

\begin{figure}
	\centering
	\includegraphics[width=0.95\columnwidth]{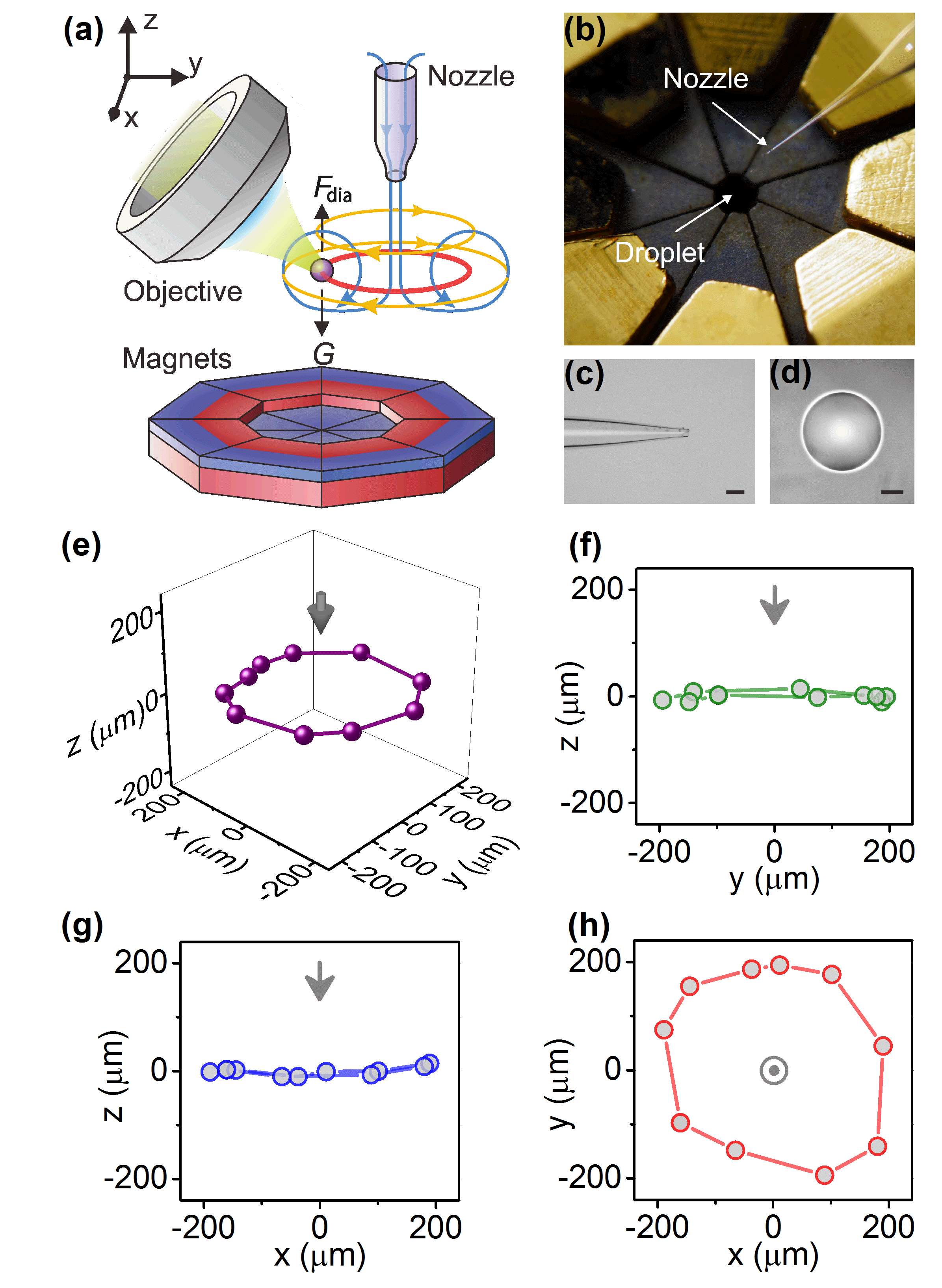}
	\caption{Experimental trap of droplet using hydrodynamic tweezers.
		(a) Schematic plot of construction of the hydrodynamic tweezers. An array of magnets generate the diamagnetism force to offset the gravity of droplet, then the vortex field can easily trap the water droplet. A microscope is used to record the position as well as the motion of the droplet.
		(b) Photograph of the tweezers structure. A micro water droplet can be trapped in the center region of the magnets array.
		(c) Photograph of the nozzle made from a glass tube with $1$ mm inner diameter.
		(d) Photograph of a typical micro-droplet being trapped. The droplet is spinning, which can be observed in condensing process (see Supplementary Video 1 \cite{supplementary}).
		(e) The measured positions of the micro-droplet which are obtained by slightly changing the relative location of the magnetic field to nozzle. The positions clearly show a ring structure as expected, the slight deformation is due to the imperfection of the nozzle and the boundary effect from the center region of the magnets array. Gray arrow indicates the position of the nozzle tip. (f)-(h) The projection of data in (e) to the
		$y$-$z$, $x$-$z$ and $x$-$y$ planes.
	}
\end{figure} 

To create a droplet in the fluid flow, we use a piezo setup to generate a mist of water with diameter less than $5$ $\mu$m. These small droplets spiral into the trapping ring to condense into a bigger one, the trajectories can be seen in Supplementary Video 1 \cite{supplementary}. We control the gas flow velocity at the nozzle outlet by setting the applied pressure $P$ less than 1 psi and tuning the dosing valve precisely. Then we measure the flow velocity to get the $\omega$ between $2\pi\times5$ to $2\pi\times50$ Hz. At ambient conditions, the viscosity of nitrogen gas is $17.6$ $\mu$Pas. And the droplet is liquid water of density $\rho=1.0$ g cm$^{-3}$ and  volume magnetic susceptibility $\chi_v=−9.0\times10^{-6}$. During the process of capturing these sub-micron particles with our hydrodynamic tweezers, we can get control of the droplet size by tuning the time duration of water mist generation. Droplets with diameter ranging from as large as $300$ $\mu$m to less than $10$ $\mu$m can be stably trapped. While for smaller droplet (less than $3$ $\mu$m), trapping becomes unstable and escaping becomes frequent. The reason is that, for small droplet, the instability of the flow become so significant that the drag force pulls the object along the velocity of fluid and eventually makes it fly out of the bound field as a result of centrifugal force.

 The micro-nozzle used to generate gas flow is made of a glass tube with inner diameter $1$ mm, and is fabricated by heating-pulling process. The nozzle with outlet diameter $d$ being about $20$ $\mu$m can generate a trapping ring with a radius around $210$ $\mu $m, which is insensitive to the flow velocity and droplet size. The nozzle is mounted on a manual positioner so that the position of the gas flow field can be tuned. 
 
 We build the diamagnetic trap using an array of $16$ pieces of micro-machined NdFeB permanent magnets of the same structure (see Sec. \uppercase\expandafter{\romannumeral2} of the supplemental material \cite{supplementary}), and they are hold by a copper unit. The strength of the restoring force near the equilibrium is $k=4\pi a^3 k_0/3$, with $k_0=4.69\times10^{7}$ Nm$^{-4}$. Theoretically, the droplet is steadily trapped at any position along the trapping ring, but in the presence of the diamagnetism-gravity potential, there is only one trapping point close to the center of magnetic bound field. The position of the magnets can be slightly tuned to generate an external force $\bm{F}_{\textrm{ext}} $, which can be utilized to pull the droplet. So the equilibrium position of droplet along the trapping ring can be controlled.
Fig.~3(e) shows the measured positions where the droplet is trapped by the tweezers. As expected, the stable equilibrium positions only exist on the trapping ring.

\subsection{Transport}

Having achieved the trapping of droplet, we continue to test transport behavior of the micro-droplet along trapping ring, the experimental process is shown in Fig.~4(a). A glass needle that works as a disturber is put to be close to the path of droplet. The van der Waals (vdW) interaction between droplet and needle surface is so strong that the droplet will attach to the surface of the needle in the absence of gas flow. For the hydrodynamic tweezers here, however, the droplet is bounded on the trapping ring. So when the external disturber is approaching, besides vdW interaction, there is also hydrodynamic interaction between the droplet and surface. Then the droplet would bypass the disturber.

The fluid flow in the vicinity of the disturber develops boundary layers. In the boundary layer, the velocity of gas flow drops down to zero when approaching the disturber surface. This changes the trapping field regularly. On the other hand, the robustness of vortex field helps keep its topological structure that remains as a closed loop even when the disturber is approaching. This ensures that the trapping ring also keeps its topological structure.

 Numerical simulations affirm that, when the needle comes close from the outer or inner sides of the trapping ring, the ring is always pushed away from the disturber but remains as a closed loop. Figs.~4(b)-4(e) plot the numerical simulations of the velocity and the vorticity fields, where the droplet can still be trapped in the ring under influence of the disturber. As shown in Figs.~4(b, c) and Figs.~4(d, e), the vortex field lines together with the trapping ring can always bypass the disturber from outer and inner sides of the ring, respectively. Figs.~4(b) and 4(d) show that, the trapping point is always pushed away from the disturber in both situations. From one perspective, the droplet feels a repulsive force from the disturber.

\begin{figure}
	\centering
	\includegraphics[width=0.95\columnwidth]{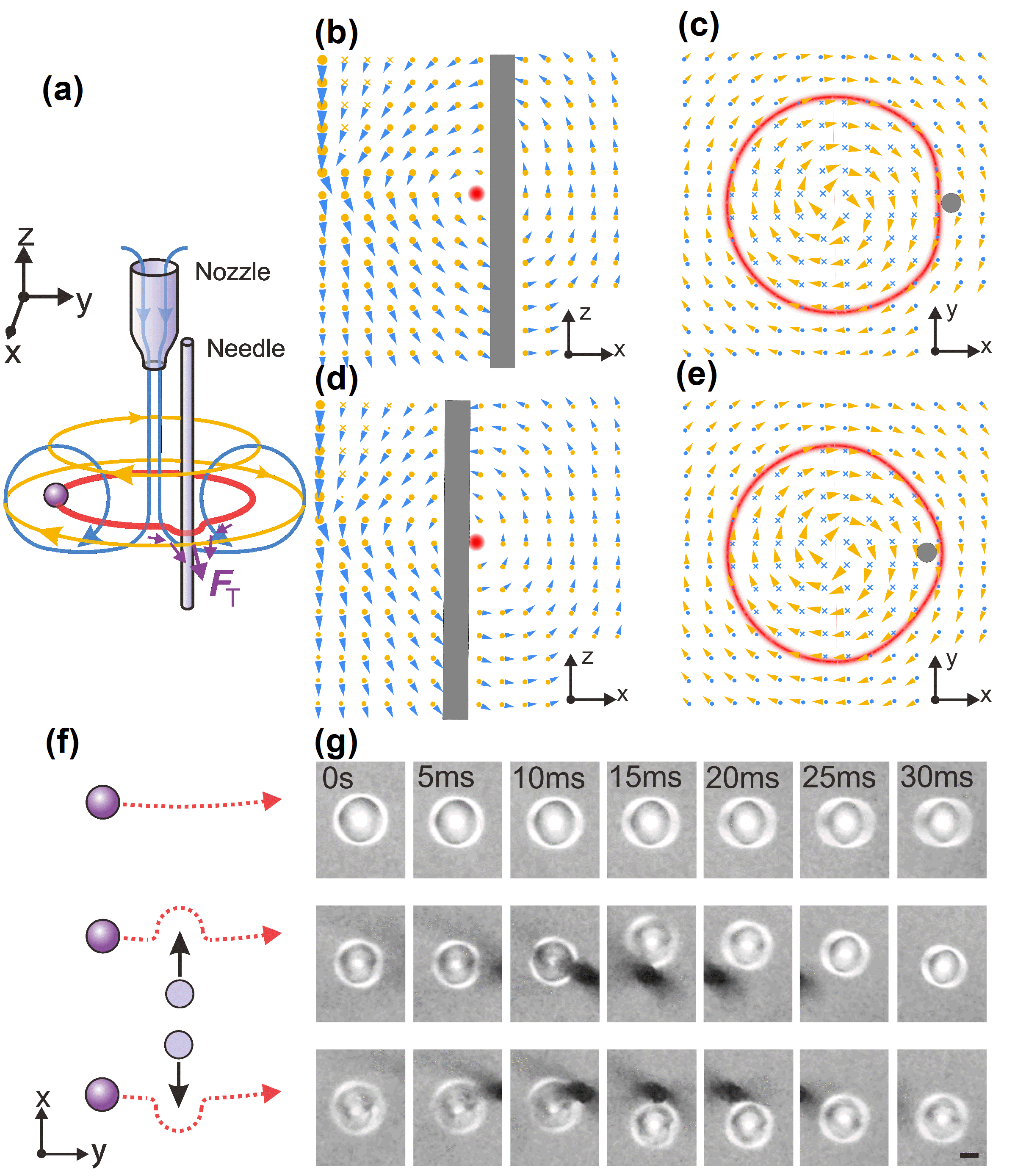}
	\caption{ Self-protected transport in trapping ring.
		(a) Schematic plot of the transport. A needle serving as an external disturber is placed into the flow field. The micro-droplet moves along the trapping ring and automatically bypasses the needle when encountering it.
		(b,~c) Numerical simulation of the velocity (blue) and vorticity (yellow) fields in the cross sections in $x$-$z$ and $x$-$y$ planes in (a). The disturber is approaching the trapping ring from the outer side. The size of arrows is proportional to the field strength. The red dot in (b) (red ring in (c)) indicates the trapping point (trapping ring).
		(d,~e) Same as (b,~c), but with disturber approaching the trapping ring from the inner side.
		(f) Scheme of three experiments processes. Top to bottom: without a needle, and with needle coming close from the outer and inner sides of the trapping ring. 
		(g) Time-series photographs of transport processes corresponding to (f), with the scale bar being $20$ $\mu$m (see Supplementary Video 4 and 5 \cite{supplementary}).
	}
	\label{simulation}
\end{figure}

In order to check the proposed mechanism, an external force $\bm{F}_{\textrm{ext}}$ is generated by tuning the position of external magnetic field and a glass needle with diameter $20$ $\mu$m is used as a disturber. We can pull the droplet along trapping ring with or without a disturber. When there is no disturber, the droplet can travel along the ring smoothly (see Supplementary Video 2 \cite{supplementary}). When the disturber is approaching in all directions, droplet trapped in the ring always feels repulsive force (see Supplementary Video 3 \cite{supplementary}). It is also observed that, no matter where to put the disturber, the droplet would always bypass the needle, as shown in Figs.~4(f) and 4(g)  (see Supplementary Video 4 and 5 \cite{supplementary}). For comparison, a suspended droplet without the presence of gas flow is attracted to the needle (see Supplementary Video 6 \cite{supplementary}). Hence, the hydrodynamic tweezers protect the droplet from being caught by external objects.  Such a protection mechanism is, on one hand, from the overall topological property of the vortex field which cannot be changed by local disturbance, and on the other hand, from the robust reconstruction of trapping force field which automatically pushes the trapping ring away from the disturber.

\begin{figure}
	\centering
	\includegraphics[width=0.95\columnwidth]{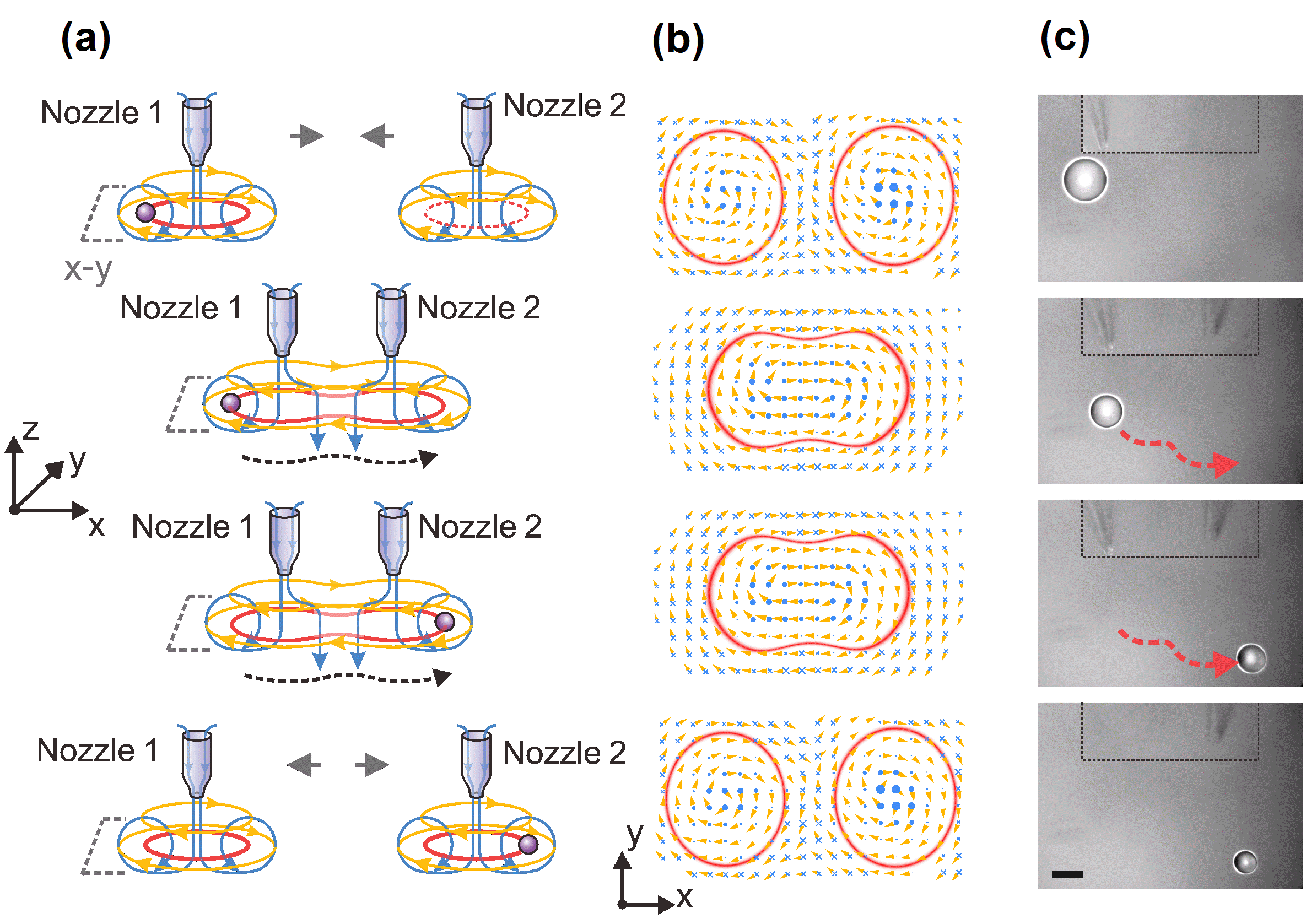}
	\caption{Topological transfer between two tweezers.
		(a) Schematic plot of transfer of the droplet from one trapping ring to the other. Top to bottom: the process begins with two tweezers far away from each other, and a droplet being trapped in the left one; then, we move the tweezers close to each other, so that the topological structure of the vortex field is transformed from two separated loops into a single deformed one; next, we conduct the process of transferring the droplet from left to right using a small external force; finally, the two nozzles are moved away from each other and the topological structure of trapping ring returns to the separated case with the droplet being trapped in the right tweezers.
		(b) Numerical simulations of the velocity (blue) and vorticity (yellow) fields in the cross sections in $x$-$y$ plane in (a). 
		(c) Photograph of droplet in the transfer process described by (a). Inset is the photograph of nozzle tips (see Supplementary Video 7 \cite{supplementary}).
	}
	\label{field}
\end{figure}


\subsection{Transfer}

 Furthermore, we explore the transfer process of the droplet between two hydrodynamic tweezers. As described above, a trapping ring exists as a closed loop that does not break. So in the presence of the second tweezers, there are only two topologies, one is two independent trapping rings and the other is a single loop that is created by merging two rings. With such topologies, we achieve the transfer of a droplet from one tweezers to the other, as schematically shown in Fig.~5(a). This transfer process is confirmed with numerical simulation shown in Fig.~5(b). We have also performed the experiments, and photographs of the four steps in a typical transfer process are presented in Fig.~5(c) (see also Supplementary Video 7 \cite{supplementary}). After many tries of the transfer processes, we find that the process is not sensitive to the position of nozzle, the relative angle of the two nozzles and the size of the trapping ring, which is readily achieved as a character of topology.


\section{CONCLUSION}
In summary, we have explored and realized a non-contact method to manipulate micro-objects in an easy-controllable way and have shown that the topological property of fluid flow can be utilized for practical applications. The special character of the hydrodynamic tweezers is that the target object is protected from external disturbances, such as avoiding attaching to a surface. And with an adjustable trapping force, the micro-object manipulation techniques demonstrated here will be useful in those scenarios where strong interactions are involved like scanning microscopy \cite{optic_tweezer_scanning}, short-range weak interaction sensing \cite{short_range, gravity}, as well as a broad range of applications that involve rotation degrees of freedom \cite{opto-mecha}.

\begin{acknowledgments}

	We thank Professor Xiaoshun Jiang in Nanjing University for discussions. This work was supported by National Natural Science Foundation of China (Grant Nos. 81788101, 11675163 and 11227901), the National Key Basic Research Program of China (Grant No. 2013CB921800), the CAS (Grant No. QYZDY-SSW-SLH004 and No. GJJSTD20170001) and the Anhui Initiative in Quantum Information Technologies (Grant No. AHY050000).
	
\end{acknowledgments}


\end{document}


\title{Supplemental Information for \\Manipulation of a Micro-object  Using Topological Hydrodynamic Tweezers}


\author{Peiran Yin}
\thanks{These authors contributed equally to this paper and are joint first
	authors.}
\affiliation{CAS Key Laboratory of Microscale Magnetic Resonance and Department of Modern Physics, University of Science and Technology of China, Hefei 230026, China}
\affiliation{Synergetic Innovation Center of Quantum Information and Quantum Physics, University of Science and Technology of China, Hefei 230026, China}
\author{Rui Li}
\thanks{These authors contributed equally to this paper and are joint first
	authors.}
\affiliation{CAS Key Laboratory of Microscale Magnetic Resonance and Department of Modern Physics, University of Science and Technology of China, Hefei 230026, China}
\affiliation{Synergetic Innovation Center of Quantum Information and Quantum Physics, University of Science and Technology of China, Hefei 230026, China}
\affiliation{Hefei National Laboratory for Physical Sciences at the Microscale, University of Science and Technology of China, Hefei 230026, China}
\author{Zizhe Wang}
\author{Shaochun Lin}
\author{Tian Tian}
\author{Liang Zhang}
\author{Longhao Wu}
\author{Jie Zhao}
\author{Changkui Duan}
\affiliation{CAS Key Laboratory of Microscale Magnetic Resonance and Department of Modern Physics, University of Science and Technology of China, Hefei 230026, China}
\affiliation{Synergetic Innovation Center of Quantum Information and Quantum Physics, University of Science and Technology of China, Hefei 230026, China}
\author{Pu Huang}
\thanks{Corresponding author: \href{mailto:hp@nju.edu.cn}{hp@nju.edu.cn}}
\affiliation{National Laboratory of Solid State Microstructures and Department of Physics, Nanjing University, Nanjing, 210093, China}
\author{Jiangfeng Du}
\thanks{Corresponding author: \href{mailto:djf@ustc.edu.cn}{djf@ustc.edu.cn}}
\affiliation{CAS Key Laboratory of Microscale Magnetic Resonance and Department of Modern Physics, University of Science and Technology of China, Hefei 230026, China}
\affiliation{Synergetic Innovation Center of Quantum Information and Quantum Physics, University of Science and Technology of China, Hefei 230026, China}
\affiliation{Hefei National Laboratory for Physical Sciences at the Microscale, University of Science and Technology of China, Hefei 230026, China}

\maketitle

\newcommand{\LJ}[1]{#1}
\setcounter{figure}{0}
\renewcommand{\thefigure}{S\arabic{figure}}

\makeatother

\global\long\def\s{\sigma}
 \global\long\def\k{\kappa}
 \global\long\def\ph{\hat{n}}
 \global\long\def\aa{\hat{a}}
 \global\long\def\bra#1{\left\langle #1\right|}
 \global\long\def\ket#1{\left|#1\right\rangle }


\baselineskip24pt


\pagebreak[4]

\section{ Equations of the fluid field and numerical simulations}
To calculate the fluid field, we employed the Navier-Stokes momentum equation of a subsonic compressible fluid \cite{ns-equ}:
\begin{equation}
\begin{split}
\label{momen_cl}
\rho(\frac{\partial \bm{\bm{\upsilon}} }{\partial t}+\bm{\bm{\upsilon}}\cdot \bigtriangledown \bm{\bm{\upsilon}})=-\bigtriangledown \bm{\bar p}+\mu \bigtriangledown ^2\bm{\bm{\upsilon}}+\frac{1}{3}\mu \bigtriangledown (\bigtriangledown \cdot \bm{\bm{\upsilon}}),
\end{split}
\end{equation}
where $\rho$, $\bm{\bm{\upsilon}}$, $t$, $\bm{\bar p}$ and $\mu$ are the density, flow velocity, time, dynamic pressure and dynamic viscosity, respectively.
In addition, there is the mass continuity equation:
\begin{equation}
\begin{split}
\label{mass_cl}
\frac{\partial\rho}{\partial t}+\bigtriangledown \cdot (\rho \bm{\bm{\upsilon}})=0,\\
\end{split}
\end{equation}
and the energy conservation equation:
\begin{equation}
\begin{split}
\label{ener_cl}
\frac{\partial (\rho E)}{\partial t}+\bigtriangledown \cdot [\bm{\bm{\upsilon}}(\rho E+p) ]=\bigtriangledown \cdot [k_{\textrm e}\bigtriangledown T-h\bm{J}+(\tau_{\textrm e}\cdot \bm{\bm{\upsilon}})].
\end{split}
\end{equation}
In the last equation, $E$ is the total energy of the fluid, including internal energy and kinetic energy. $T$ is the temperature,  $k_{\textrm e}$ is the effective coefficient of heat conduction, $\tau_{\textrm e}$ is the effective deviatoric stress tensor, $h$ is the enthalpy of the gas, and $\bm{J}$ is the diffusion flux.

We conducted numerical computations by employing the commercial CFD software ANSYS FLUENT, which simulates the subsonic incompressible gas flow field with the Laminar viscous simulation model.  Because the $Re$ number in our experiment is so small, we treated the gas flow as incompressible flow. We enabled viscous heating option. So the Energy equation option is also activated, and we chose the gas density as constant in the material module. The computation domain is a cylinder with  a half of a nozzle dipped inside. The cylinder's height is $1200~\mu$m, and its radius is $600~\mu$m. This domain is composed of around $2\times 10^6$ mixed cells including tetrahedron cells and hexahedron cells. The amount of cells was chosen on the basis of convergence analysis using $x,~y,~ z$ velocities and continuity condition. The mesh was refined near the model surface to provide $y^+\approx 1$, so as to resolve viscous layers more accurately.  For boundary conditions, the velocity inlet boundary condition was employed. 
We set the inlet velocity around 0.05 $m/s$ to be consistent with experimental results. The pressure outlet boundary condition was used. The Gauge pressure was set as $1.0\times 10^5$ Pa. Average pressure specification and non-reflecting boundary options were activated. No-slip boundary condition was applied on the solid walls, the surface of droplet and disturber.

In the simulation of double nozzles condition, nothing is different from the single one except that we add another nozzle with the same size parallel to the original one. By changing distance between the two nozzles which is in the $x$ direction, we can get different fluid fields.

\section{  Designing of the magnets for diamagnetism levitation }
The design of the permanent magnets array was also assisted by numerical calculations based on finite element simulations. The diamagnetism force of an object in magnetic field can be expressed as \cite{diamagnetism}
\begin{align}
\bm{F}_{\rm dia}(\bm{r}) = -\frac{|\chi| V}{\mu_0} [\nabla \cdot \bm{B}(\bm{r})] \bm{B}(\bm{r}).
\end{align}
Here $\chi$ is the magnetic susceptibility, $V$ is the volume of the droplet, $\mu_0$ is the permeability of free space and $ \bm{B}(\bm{r})$ is the magnetic field vector. According to the fundamental property of magnetic field, it is impossible to generate stable trapping according to Earnshaw's theorem. However, in the presence of gravity, such stable levitation is achieved.  The designing of magnets is similar to that reported previously \cite{magnets}, with numerical simulations to optimize the structure and the geometry of the permanent magnets. In our designing, 16 pieces of micro-machined NdFeB permanent magnets of the same structure were used to construct the double-decker magnets set. The geometry of the magnets was optimized to offset the gravity of water droplet effectively. Its  schematic diagram is shown in Fig.~S1(a). The upper eight slices of magnets are magnetized along the radial direction, while the other eight slices of magnets in the lower layer are magnetized in the opposite direction. The magnets set is hold by a copper structure. 

\begin{figure}
	\centering
	\includegraphics[width=0.95\textwidth]{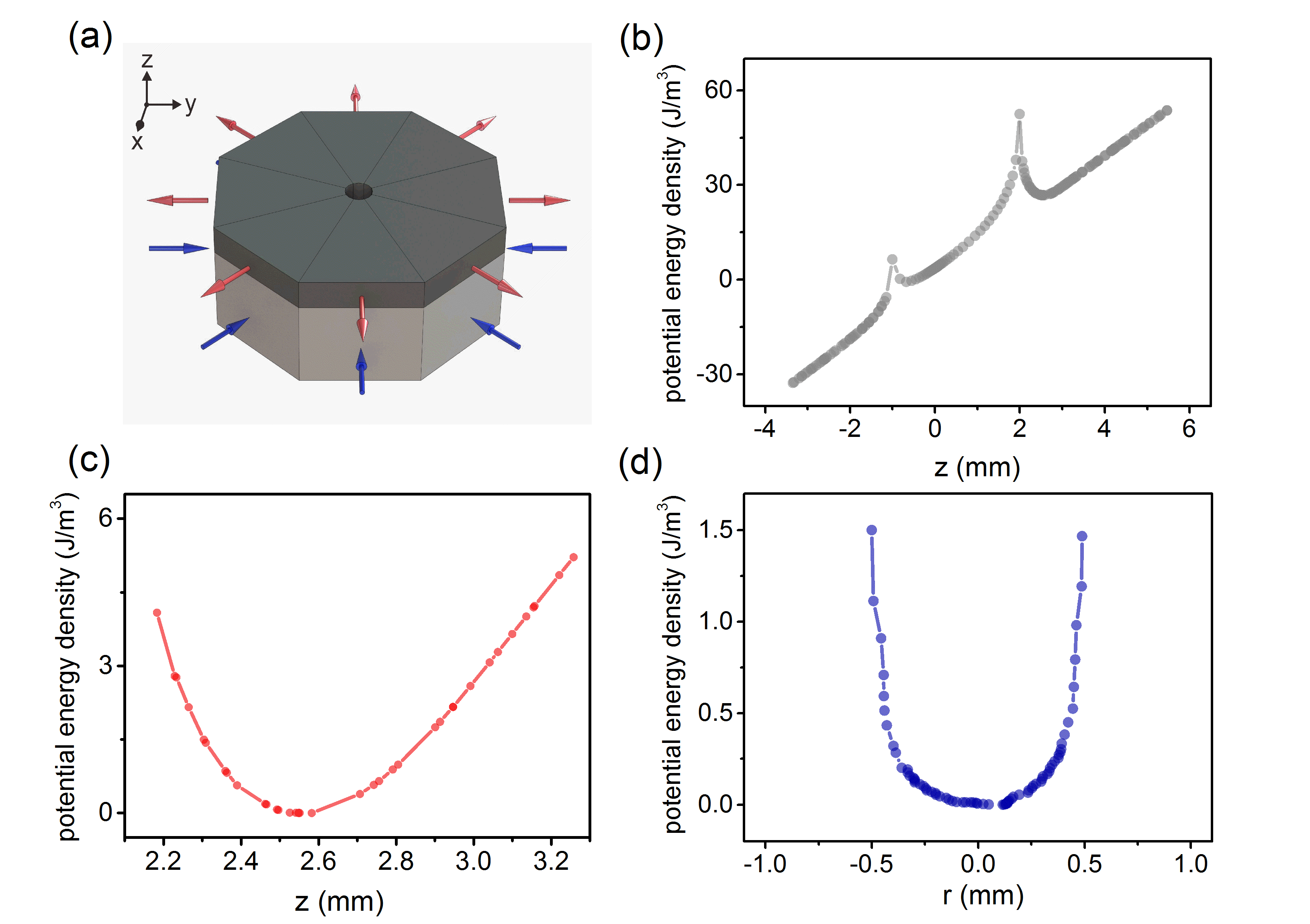}
	\caption{Design of the magnets for diamagnetism levitation.
		(a) Schematic diagram of the magnets set containing two layer of magnets with opposite magnetization.
		(b) $z$-direction distribution of the total potential energy density of pure water as a sum of the magnetic energy and gravitational potential energy.
		(c) Zooming in around the upper singular point in (b).
		(d) Radial-direction distribution of the potential energy density.
	}
\end{figure}

The magnetic-gravity trap can be described as a harmonic potential near the equilibrium position, i.e., $U(\bm{r}) \approx \frac{1}{2} k z^2 + \frac{1}{2} b x^2+ \frac{1}{2} b y^2$. In the region where typical gas flow exists, the restoring force is $\bm{F}(\bm{r}) = - \nabla U(\bm{r})$. So the restoring force in the $z$ direction near the equilibrium position can be approximatively expressed as $\bm{F}_\textrm{m}=-k\bm{z}$. And the external force $\bm{F}_{\textrm ext} $ used to pull the droplet can be generated using this diamagnetism force by relocating the magnets by a distance $\delta \bm{r} $, and as a result, $\bm{F}_{\textrm ext}= - (\delta \bm{r} \cdot \nabla) \nabla U(\bm{r}) $. Figs.~S1(b)-(d) plot the total potential energy density of a small water droplet in the magnetic field. In the vertical axis ($z$-axis), two singular points exist and the upper one which is stable and optically accessible is employed in our experiment.

To load the droplet into the tweezers, deionized water is first atomized using a home-built piezoelectric dispenser close to the tweezers region. The water mist is attracted into the tweezers from the dispenser and condenses into a single large droplet, whose size can be readily regulated by controlling the duration of atomization. As the experiments are carried out at room temperature in  nitrogen gas, the droplet evaporates gradually after the atomization process. A droplet with a diameter of $100$ $\mu$m typically has a life span less than $30$\ s, which depends on environment condition such as the humidity. And the evaporation speeds up as the droplet size decreases. Measurements are performed using a microscope, and a charge-coupled device (CCD) camera with maximum speed of $1000$ frames per second was used to record the position and motion information.